\documentclass{ws-procs9x6}

\begin{document}

\title{Opening the Window for Technicolor
}

\author{Deog Ki Hong\footnote{\uppercase{O}n leave from
\uppercase{D}epartment of \uppercase{p}hysics,
\uppercase{p}usan \uppercase{n}ational
\uppercase{u}niversity, \uppercase{p}usan 609-735,
\uppercase{k}orea}}

\address{Center for Theoretical Physics,\\
Massachusetts Institute of Technology, \\
Cambridge, MA 02139, USA\\
E-mail: dkhong@pusan.ac.kr, dkhong@lns.mit.edu}

\author{Stephen D.~H. Hsu}

\address{Department of Physics,\\
University of Oregon, \\
Eugene OR 97403-5203, U.S.A.\\
E-mail: hsu@duende.uoregon.edu}

\author{Francesco Sannino}

\address{The Niels Bohr Institute,\\
Blegdamsvej  17, DK-2100 Copenhagen \O,  Denmark \\
E-mail: francesco.sannino@nbi.dk}


\maketitle
\vskip -4.2in
\hfill{\parbox[b]{1in}
{\hbox{\bf\tt PNUTP-04/A10, MITCTP-3552}}}

\vskip 4.2in

\abstracts{
Recently a new class of technicolor models are proposed, using
technifermions of symmetric second-rank tensor.
In the models, one can make reasonable
estimates of physical quantities like the Higgs mass and the size of
oblique corrections, using
a correspondence to super Yang-Mills theory in the Corrigan-Ramond
limit. The models predict a surprisingly light Higgs of mass,
$m_H=150\sim\,500\,{\rm GeV}$ and have naturally small $S$
parameter.}

The standard model for the interaction of elementary particles
has so far passed all experimental tests. Its gauge structure,
$SU(3)_c\times SU(2)_L\times U(1)_Y$ is extremely well tested
and its flavor structure is  measured precisely.
Hence, we now firmly believe that the standard model is the correct
theory for elementary particles at the shortest distance we have ever
explored, though the Higgs, introduced in the standard model
to account for the electroweak symmetry breaking, is yet to be found.


The Higgs, which is the only undiscovered particle in the standard model,
poses therefore pressing challenges for both theorists and experimentalists.
As a scalar particle, its mass is
extremely sensitive
to the scale of new physics or the ultraviolet cutoff, $\Lambda$,
of the standard model.
By naive dimensional analysis the Higgs mass is given as
\begin{equation}
m^2_H=c\,\Lambda^2,
\end{equation}
where the dimensionless constant $c$ is $O(1)$ and
depends logarithmically on $\Lambda$.
If the scale of new physics is higher than $100~{\rm TeV}$,
it requires a severe fine-tuning, $c\lesssim 10^{-4}$ to get a light Higgs,
$m_H<1~{\rm TeV}$, as needed for perturbative unitarity of the standard model.

The fine-tuning problem associated with the Higgs, known as the
hierarchy  problem, has been one of the most
fundamental problems in particle physics since the seventies.
The earliest attempt~\cite{Susskind:1978ms}
to solve the hierarchy problem was to introduce a
new strong interaction, Technicolor, that breaks the electroweak symmetry
dynamically at $\Lambda_{\rm TC}\sim 1\,{\rm TeV}$
where the new interaction becomes strong.
The Higgs is then a composite particle, made of strongly bound
technifermions.
Another attempt was made soon after by supersymmetrizing
the standard model\cite{Dimopoulos:1981zb}.
The supersymmetric extension of the
standard model was accepted quickly due to the fact that
it is perturbative, consistent with experimental data~\cite{Ellis:2004aq}, and
furthermore it indicates gauge coupling
unification~\cite{Ellis:1990zq}.
(See for a drastically different view on the fine-tuning
problem~\cite{Arkani-Hamed:2004fb}.)

On the other hand, technicolor models have been largely abandoned,
though several interesting ideas were introduced in recent years~\cite{Hill:2002ap}.
It is often claimed in the literature
that  technicolor is ruled out by electroweak precision
data. However, the real killer of technicolor is not the electroweak precision
data but rather our ignorance of strong dynamics and thus inability
to make a systematic and precise estimate of physical quantities like the
Higgs mass or the oblique corrections to the electroweak observables.
Lacking a reliable means to solve strongly interacting systems,
analyses have been made in analogy with QCD,
and use the experimental hadronic data to study
technicolor models.
For instance, if one uses for technicolor a scaled-up version
of QCD, the $S$-parameter of the oblique
corrections~\cite{Kennedy:1988sn} will be
given as~\cite{Peskin:1991sw}
\begin{equation}
S\approx 0.11\,N_{TC}\,N_D\,,
\end{equation}
where $N_{TC}$ is the number of technicolors and $N_D$ is the
number of $SU(2)_L$ technifermion doublets.
Since experimentally the $S$ parameter from physics beyond standard
model~\cite{Eidelman:2004wy}
is measured to be $S=-0.13\,\pm\,0.10$,
the QCD-like technicolor model is roughly $(N_{TC}\,N_D+1)\,\sigma$
away from the electroweak
precision data, indicating that technicolor models with
many electroweak doublets are presumably ruled out.

However, the simple QCD-like technicolor models are already ruled out
by the constraint on the flavor-changing neutral currents when one tries
to explain the fermion masses with additional strong
interactions (ETC)~\cite{Dimopoulos:1979es}.
The smallness of the flavor-changing neutral currents as in
the mass difference in $K_L$ and $K_S$ requires the
ETC scale to be larger than $10^3~{\rm TeV}$ for QCD-like models,
which would then lead to too small fermion masses.
To solve this problem, technicolor models with a quasi UV-fixed point,
called walking technicolor~\cite{Yamawaki:1985zg},
were suggested, where technifermion bilinear operators have
anomalous dimension, $\gamma_m\simeq1$, allowing  lower ETC scales
$\Lambda_{\rm ETC}\,\sim\, 100\,{\rm TeV}$.
Furthermore, contributions to the $S$ parameter will be somewhat reduced due to
walking~\cite{Sundrum:1991rf}.

The $\beta$ function may be expanded in powers of coupling $\alpha$.
\begin{eqnarray}
\beta(\alpha)\equiv\mu\frac{\partial}{\partial\mu}\alpha
=-b\,\alpha^2-c\,\alpha^3+\cdots\nonumber
\end{eqnarray}
We need $b>0$ to be asymptotically free and $c<0$  for a UV fixed
point. Then UV fixed point is approximately
$\alpha_*\simeq-{b}/{c}$. To break electroweak symmetry, the
critical coupling for chiral symmetry breaking, which in the
ladder approximation is $\alpha_c\simeq{\pi}/(3C_2(R))$, should be
smaller than the coupling at the UV fixed point. ($C_2(R)$ is the
eigenvalue of the quadratic Casimir operator for fermions in a
representation $R$.)
For a $SU(N)$ technicolor gauge theory with $N_f$ technifermions
we have listed $b$, $c$, and the critical couplings for the chiral
symmetry breaking in Table~1. We see that the critical number of flavors
for a $SU(N)$ technicolor gauge theory
to have a walking coupling constant is $N_f\sim 4N$
if technifermions are in
the fundamental representation, while $N_f=2$ if
in  the second rank symmetric tensor representation
(See Table~1).
The most economic way to have a walking coupling is therefore to introduce
a technifermion doublet in the second-rank symmetric tensor
representation or S-type
for $SU(N)_{\rm TC}$ with $N\le5$~\cite{Sannino:2004qp}.
\begin{table}[t]
\tbl{$SU(N)_{TC}$ technicolor with $N_f$ technifermions in either
 fundamental representation or in second rank tensor representation.
 The upper sign is for the symmetric tensor and the lower sign is for
 the antisymmetric tensor.}
{
\begin{tabular}{|c|c|c|c|}
\hline
     & { fundamental representation} &
     { (anti-)symmetric second rank  tensor} \\ \hline
$6\pi \,b$  & $11N-2N_f$ & $11N-2N_f (N\pm2)$   \\ \hline
$24\pi^2\,c$  &$34N^2-10N\,N_f$ &
$34N^2-10N\,N_f\,(N\pm2)$ \\
  &$-3(N-{1}/{N})N_f$ &
$-{6}(1\mp1/N)(N\pm2)^2N_f\,$ \\ \hline
$c<0$ & $N_f>{34N^2}/({13N-3/N})$ & $N_f(N\pm2)>{34N^2}/
({16N\pm6-12/N})$ \\ \hline
$\alpha_c$  & ${2\pi N}/({3N^2-3})\, $& ${2\pi N}/
[3{(N\pm2)(N\mp1)}] $  \\ \hline
$N_f^{\rm crit}$  & $N_f\sim4N$&  $N_f=2$ for S-type and $N\le5$
 \\ \hline
\end{tabular}
\label{table1} }
\end{table}

Unlike previous technicolor models, new
technicolor models~\cite{Sannino:2004qp,Hong:2004td}, where technifermions
are in the second-rank symmetric tensor representation,
allow systematic estimates of the Higgs mass~\cite{Hong:2004td}
and other physical observables~\cite{hong2}.
This is possible because in the large $N$ Corrigan-Ramond
limit~\cite{Corrigan:1979xf},
it is mapped into super Yang-Mills when $N_f=1$~\cite{Armoni:2004uu}.
Not only one can export the exact results established in super
Yang-Mills to nonsupersymmetric theories by considering 1/N
corrections \cite{Sannino:2003xe} but one can also make relevant
predictions about, previously unknown, nonperturbative aspects of
super Yang-Mills \cite{Feo:2004mr}.

In the large $N$ limit, the Higgs particle is identified with the
scalar fermion-antifermion state whose pseudoscalar partner in
ordinary QCD is the $\eta^{\prime}$.  The low lying bosonic sector
contains precisely a scalar and a pseudoscalar meson.
In the supersymmetric limit we can
relate the masses to the fermion condensate $\left< \widetilde{q}q
\right> \equiv \left< \widetilde{q}^{\{i,j \}} {q}_{\{i,j \}}
\right> $ \cite{Sannino:2003xe}:
\begin{eqnarray}
M=\frac{2\,\alpha}{3}\, \left[\frac{3 \left< \widetilde{q}q
\right> }{32\pi^2\,N} \right]^{\frac{1}{3}} =
\frac{2\hat{\alpha}}{3} \Lambda \ ,
\end{eqnarray}
with $\left< \widetilde{q}q \right> = 3N\Lambda^3$ and $\Lambda$
the 
invariant scale of the theory:
\begin{eqnarray}
\Lambda^3 = \mu^3 \left(\frac{16\pi^2}{3Ng^2(\mu)}\right) \exp
\left[ \frac{-8\pi^2}{Ng^2(\mu^2)}\right] \ .
\end{eqnarray}
We have also defined $\hat{\alpha} = \alpha\,[9/(32\pi^2)]^{1/3}$.
The unknown $O(1)$ numerical parameter $\hat{\alpha}$
is the coefficient of the K\"{a}hler term in the
Veneziano-Yankielowicz effective Lagrangian describing the lowest
composite chiral superfield. By analogy with QCD, we take
$\hat{\alpha} \sim1-3$.  Then, the Higgs mass is estimated as
$m_{H}=M\simeq 150-500~{\rm GeV} \ .$
Here we have chosen $\Lambda =\Lambda_{TC}\sim 250~$GeV.
The $1/N$ corrections to the Higgs mass can be made
systematically~\cite{Sannino:2003xe,Hong:2004td}.
Similarly, the oblique corrections are calculated
precisely in the large $N$ limit~\cite{hong2}.

In conclusion, the newly proposed technicolor models
have overcome the typical barrier of technicolor
theory, which is the inability to calculate precisely physical
quantities like the Higgs mass or $S$ parameters.
This is possible due to their correspondence to Super Yang-Mill
theories in the large $N_{TC}$ limit.
Somewhat surprisingly the new technicolor models  naturally
produce light composite Higgs bosons. The models are also nearly
conformal with the minimal number of flavors, making the $S$ parameter
naturally small.

\section*{Acknowledgments}
The work of D.K.H. is supported by Korea Research Foundation Grant
(KRF-2003-041-C00073) and also in part by
funds provided by the U.S. Department of Energy (D.O.E.)
under cooperative research agreement \#DF-FC02-94ER40818.
The work of S.H. was supported in part
under DOE contract DE-FG06-85ER40224.

\end{document}